%% file: DiBoPhotons_PLB.tex
\documentclass[preprint,number,sort&compress,twocolumn,3p]{elsarticle}
\pdfoutput=1
\usepackage{amsmath,amsthm,amssymb,multirow,psfrag}
\usepackage{epsfig}
\usepackage{color}
\usepackage[utf8]{inputenc} 

\graphicspath{{./Figures/}}

\begin{document}

\def\lsim{\mathrel{\rlap{\lower4pt\hbox{\hskip1pt$\sim$}}
  \raise1pt\hbox{$<$}}}
\def\gsim{\mathrel{\rlap{\lower4pt\hbox{\hskip1pt$\sim$}}
  \raise1pt\hbox{$>$}}}
\newcommand{\vev}[1]{ \left\langle {#1} \right\rangle }
\newcommand{\bra}[1]{ \langle {#1} | }
\newcommand{\ket}[1]{ | {#1} \rangle }
\newcommand{\ev}{ {\rm eV} }
\newcommand{\kev}{{\rm keV}}
\newcommand{\mev}{{\rm MeV}}
\newcommand{\tev}{{\rm TeV}}
\newcommand{\mpl}{$M_{Pl}$}
\newcommand{\mw}{$M_{W}$}
\newcommand{\Ft}{F_{T}}
\newcommand{\Zparity}{\mathbb{Z}_2}
\newcommand{\BLambda}{\boldsymbol{\lambda}}
\newcommand{\met}{\;\not\!\!\!{E}_T}
\newcommand{\be}{\begin{equation}}
\newcommand{\ee}{\end{equation}}
\newcommand{\bea}{\begin{eqnarray}}
\newcommand{\eea}{\end{eqnarray}}
\newcommand{\nn}{\nonumber}
\newcommand{\gev}{{\mathrm GeV}}
\newcommand{\hc}{\mathrm{h.c.}}
\newcommand{\eps}{\epsilon}
\newcommand{\bwt}{\begin{widetext}}
\newcommand{\ewt}{\end{widetext}}
\newcommand{\draftnote}[1]{{\bf\color{blue} #1}}

\newcommand{\cO}{{\cal O}}
\newcommand{\cL}{{\cal L}}
\newcommand{\cM}{{\cal M}}

\newcommand{\fref}[1]{Fig.~\ref{fig:#1}} 
\newcommand{\eref}[1]{Eq.~\eqref{eq:#1}} 
\newcommand{\aref}[1]{Appendix~\ref{app:#1}}
\newcommand{\sref}[1]{Section~\ref{sec:#1}}
\newcommand{\tref}[1]{Table~\ref{tab:#1}}

\title{\Large{{\bf New vector bosons and the diphoton excess}}}
\author[infn]{Jorge de Blas\corref{cor1}}
\ead{Jorge.DeBlasMateo@roma1.infn.it}

\author[ugr]{Jos\'e Santiago}
\ead{jsantiago@ugr.es}

\author[ugr]{Roberto Vega-Morales}
\ead{rvegamorales@ugr.es}

\cortext[cor1]{Corresponding author}

\address[infn]{INFN, Sezione di Roma, Piazzale A. Moro 2, I-00185 Rome, Italy}

\address[ugr]{Departamento de F\'{i}sica Te\'{o}rica y del Cosmos and CAFPE,
Universidad de Granada, Campus de Fuentenueva, E-18071 Granada, Spain}

\input{DiBoPhotons_main.tex}

\bibliographystyle{apsrev}
\bibliography{DiBoPhotons}

\end{document}

%% file: DiBoPhotons_main.tex

\begin{abstract}
We consider the possibility that the recently observed diphoton excess
at $\sim 750$~GeV can be explained by the decay of a scalar particle
($\varphi$) to photons.~If the scalar is the remnant of a
symmetry-breaking sector of some new gauge symmetry, its 
coupling to photons can be generated by loops of the charged massive vectors
of the broken symmetry.~If these new $W^\prime$ vector bosons carry
color, they can also generate an effective coupling to gluons.~In this case
the diphoton excess could be entirely explained in a simplified model
containing just $\varphi$ and $W^\prime$.~On the other hand, if $W^{\prime}$ 
does not carry color, we show that, provided
additional colored particles exist to generate the required $\varphi$
to gluon coupling, the diphoton excess could be explained by the same 
$W^{\prime}$ commonly invoked to explain the
diboson excess at $\sim 2$~TeV. We also explore possible connections
between the diphoton and diboson excesses with the anomalous $t\bar{t}$
forward-backward asymmetry.
\end{abstract}

\maketitle

\section{Introduction} \label{sec:intro} 

The successful first year of the Large Hadron Collider (LHC) Run 2 has
provided us with a relatively small amount of data at $\sqrt{s}=13$
TeV but a very interesting surprise.~Both ATLAS and CMS have reported
an excess in the diphoton spectrum with a peak in the diphoton
invariant mass at around $750$ GeV, with a statistical significance 
in the 2-4 $\sigma$ range depending on assumptions about the total width 
and the look-elsewhere effect~\cite{ATLASdiphoton,CMS:2015dxe}.~The fact that
both experiments see an excess at the same diphoton invariant mass
and
that such an excess can be compatible with the Run-1 results
has triggered an explosion of theoretical 
activity~\cite{Harigaya:2015ezk,
Mambrini:2015wyu,
Nakai:2015ptz,
Backovic:2015fnp,
Angelescu:2015uiz,
Knapen:2015dap,
Franceschini:2015kwy,
Buttazzo:2015txu,
Pilaftsis:2015ycr,
DiChiara:2015vdm,
Higaki:2015jag,
McDermott:2015sck,
Ellis:2015oso,
Low:2015qep,
Bellazzini:2015nxw,
Gupta:2015zzs,
Petersson:2015mkr,
Molinaro:2015cwg,
Dutta:2015wqh,
Cao:2015pto,
Kobakhidze:2015ldh,
Martinez:2015kmn,
Cox:2015ckc,
Becirevic:2015fmu,
No:2015bsn,
Chao:2015ttq,
Fichet:2015vvy,
Curtin:2015jcv,
Bian:2015kjt,
Chakrabortty:2015hff,
Ahmed:2015uqt,
Agrawal:2015dbf,
Csaki:2015vek,
Falkowski:2015swt,
Aloni:2015mxa,
Bai:2015nbs,
Gabrielli:2015dhk,
Benbrik:2015fyz,
Kim:2015ron,
Alves:2015jgx,
Megias:2015ory,
Carpenter:2015ucu,
Bernon:2015abk,
Chao:2015nsm,
Arun:2015ubr,
Han:2015cty,
Chang:2015bzc,
Chakraborty:2015jvs,
Ding:2015rxx,
Han:2015dlp,
Han:2015qqj,
Luo:2015yio,
Chang:2015sdy,
Bardhan:2015hcr,
Feng:2015wil,
Antipin:2015kgh,
Wang:2015kuj,
Cao:2015twy,
Huang:2015evq,
Liao:2015tow,
Heckman:2015kqk,
Dhuria:2015ufo,
Bi:2015uqd,
Kim:2015ksf,
Berthier:2015vbb,
Cho:2015nxy,
Cline:2015msi,
Bauer:2015boy,
Chala:2015cev,
Kulkarni:2015gzu,
Barducci:2015gtd
}.

Among the many different attempts, one that has proven a successful
explanation for the diphoton excess is the presence of a new scalar,
$\varphi$, 
with effective couplings to photons and gluons\footnote{See \cite{Jaeckel:2012yz} for previous studies of (pseudo-) scalars coupled to photons and gluons.}
\bea\label{eq:LphVVs}
\mathcal{L} \supset 
\frac{ \mathcal{A}_{\gamma\gamma} }{4v_\varphi} \varphi F^{\mu\nu} F_{\mu\nu} 
+ \frac{ \mathcal{A}_{gg} }{4v_\varphi} \varphi G^{\mu\nu} G_{\mu\nu} .
\eea
It has been shown~\cite{Franceschini:2015kwy} that the excess can be
successfully explained if the partial widths into photons and gluons
satisfy
\begin{equation}
\frac{\Gamma(\varphi\to \gamma \gamma)}{M_\varphi} 
\frac{\Gamma(\varphi\to gg)}{M_\varphi} 
\approx 1.1\times 10^{-6} \frac{\Gamma_\varphi}{M_\varphi},\label{gaga_gg}
\end{equation} 
where $\Gamma_\varphi$ is the total decay width. 
Although there is a slight preference in the ATLAS measurement
for a relatively large total width $\Gamma_\varphi/M_\varphi\approx 0.06$,
a much narrower particle is perfectly compatible with the published data~\cite{Falkowski:2015swt}.

In all the examples presented so far in the literature, the effective
couplings in~\eref{LphVVs} have been generated by loops of new
fermions or scalars with electric and/or color charge.~It is plausible
however that $\varphi$ is the low-energy remnant of a new TeV scale
symmetry-breaking sector of an extended gauge symmetry.~In that case the
massive vector bosons to which $\varphi$ gives mass are natural
candidates to generate the required couplings to photons and
gluons via loops.~In fact, new vector bosons are motivated by other intriguing
excesses observed at the LHC or the Tevatron, like the $\sim 2$ TeV
diboson anomaly~\cite{Aad:2015owa,Khachatryan:2014hpa,Khachatryan:2014gha}, flavor anomalies in the $B$ sector~\cite{Allanach:2015gkd}  or
the $t\bar{t}$ forward-backward asymmetry anomaly
(see~\cite{Aguilar-Saavedra:2014kpa} for a recent review).~It is
therefore natural to investigate whether new vector bosons could be
responsible for the couplings of $\varphi$ to photons and gluons
and under which circumstances such couplings can explain the observed
diphoton excess.

We initiate an investigation of this possibility via a simplified 
model containing a new vector boson that acquires its mass through the vacuum expectation 
value (vev) of a scalar $\varphi$.~We then explore the implications for decays into electroweak vector
bosons and gluons before discussing possible connections
between the diphoton and the diboson anomalies.

\section{Simplified model\label{section:simplified_model}}

We consider an extension of the Standard Model (SM) with an extra SM singlet scalar,
$\varphi$, whose vev, $v_\varphi$, 
is responsible for the spontaneous breaking of a new 
gauge symmetry.~The heavy vector bosons of the new symmetry, $W^\prime$, 
can be in arbitrary representations of the SM $SU(3)_c\times SU(2)_L \times U(1)_Y$ 
gauge group, but for simplicity in the following we will consider it to be an electroweak singlet
with hypercharge $Q\neq 0$ and in an arbitrary color
representation.\footnote{The results presented here can be
    easily generalized for the case of higher $SU(2)_L$
    representations.}
Explicit realizations of our simplified model can be obtained
generically in extensions of the SM gauge group which are broken spontaneously
by a corresponding ``Higgs'' boson. 
A representative example of an SM extension realizing our scenario
would be that of models with vector leptoquarks. Since in this case the new particles
carry both color and electric charge, they can generate by themselves the necessary
one-loop interactions with gluons and photons. Other types of models can be found, 
for instance, in~\cite{Fornal:2015boa,Fornal:2015one}, where
the SM gauge group is extended in such a way as to unify color and baryon number
and broken by the vev of a SM singlet. After the new gauge group and electroweak symmetry
are broken one is left with, in addition to the SM particles, a colored vector boson (as well as
charged vector-like fermions) which can be used to generate the necessary gluon (and
photon) effective couplings. In~\cite{Aguilar-Saavedra:2015iew} an extended gauge
group again leads to the presence of new vector bosons, in this case carrying 
electric charge but not color.

In what follows we will assume that any mixing between $\varphi$ and
the SM Higgs is small and we neglect it. Likewise, if the $W^\prime$ is a
colorless vector with $Q=\pm 1$, in our simplified model the mixing with the 
SM $W$, $\sin{\theta_{WW^\prime}}$, must be small to be consistent with electroweak precision 
data~\cite{delAguila:2010mx} and so is also neglected. 
Note that, even if the $W^\prime$ contributions to electroweak 
precision observables were to be cancelled by the effects of extra new particles, 
thus relaxing the electroweak precision limits on the $W$-$W^\prime$ mixing, a 
sizable value of $\sin{\theta_{WW^\prime}}$ would generate tree-level decays $\varphi\rightarrow W^+W^-$
and thus is constrained by the direct bounds on $W^+W^-$ production at the LHC.\footnote{Using the results from Table 1 in Ref.~\cite{Franceschini:2015kwy}, we estimate that current searches for $W^+ W^-$ resonances limit values of $\sin{\theta_{WW^\prime}}\lesssim 0.03$.}

If $W^\prime$ obtains its mass from the vev of
$\varphi$ then the $\varphi W^\prime W^\prime$ coupling can be
expressed as 
\bea\label{eq:LphVV}
\mathcal{L} \supset \frac{ 2 M^2_{W^\prime} }{v_\varphi} \varphi W^{\prime \mu} W^\prime_{\mu},
\eea
analogous to the SM $W$ boson coupling to the Higgs.~The $W^\prime$ mass is given by
\bea\label{eq:MWpr}
M_{W^\prime}=\kappa \frac{g_{W^\prime} v_{\varphi}}{2},
\eea
where $\kappa$ is a group theory factor that depends on the $\varphi$ quantum
numbers under the broken group.~For instance, if the new symmetry is an
$SU(2)$ group and $\varphi$ is part of a triplet representation we have $\kappa=\sqrt{2}$~\cite{Aguilar-Saavedra:2015iew}.~The fact that $W^\prime$ is an electroweak singlet of hypercharge $Q$
ensures the $Z$ and photon couplings to $W^\prime W^\prime$ are given by
\bea\label{eq:Vcoup}
g_{Z W^\prime W^\prime}=-e Q s_W/c_W, \quad
g_{\gamma W^\prime W^\prime}=e Q, 
\eea
with $s_W, c_W$ the sine and cosine of the weak angle.
In our simplified scenario couplings to
the SM $W$ are zero due to the trivial $SU(2)_L$ quantum numbers and the neglected $W$-$W^\prime$ mixing (whenever this mixing is allowed), but would be interesting to consider in a more complete model.

In addition to the couplings in~\eref{LphVVs}, $W^\prime$ loops will generate
the following effective couplings,
\bea\label{eq:LphVVs:extra}
\mathcal{L} \supset 
\frac{ \mathcal{A}_{\gamma Z} }{2v_\varphi} \varphi F^{\mu\nu} Z_{\mu\nu} 
+ \frac{ \mathcal{A}_{ZZ} }{4v_\varphi} \varphi Z^{\mu\nu} Z_{\mu\nu} .
\eea
Given these effective couplings, the partial decay width
for $\varphi \to VV^\prime$ can be written as
\bea
\Gamma(\varphi \to VV^\prime) =  \frac{|\mathcal{A}_{VV^\prime}|^2 M_\varphi^3}{32
  \pi (1+\delta_{V,V^\prime})v_\varphi^2}\beta_{VV^\prime} , 
\eea
where for $M_\varphi = 750$~GeV we have 
$\beta_{VV}-1 =\mathcal{O}(M_V^2/M_\varphi^2) \ll 1$ for all vector
bosons in the SM and the Kronecker delta, $\delta_{V,V^\prime}$,
accounts for the factor of $1/2$ when the
final states are identical particles.

\begin{figure}[t]
\begin{center}
\includegraphics[scale=.7]{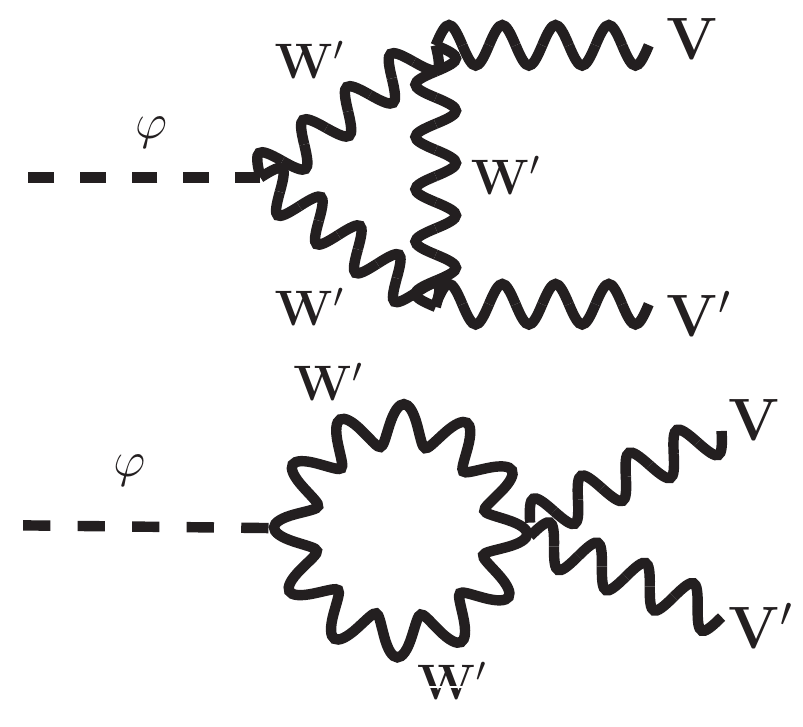}
\end{center}
\caption{$W^\prime$ contribution to the $\varphi VV^\prime$ ($V = g, \gamma, Z$) effective vertex which mediates $\varphi \to VV^\prime$ decays and production through gluon fusion.}
\label{fig:AVV_loop}
\end{figure}

Using the expression for the $W^\prime$ mass in~\eref{MWpr} we can write the partial width to mass ratios as follows 
\bea
\!\!\!\!\!\!\!\frac{\Gamma(\varphi \to VV^\prime)}{M_\varphi} =  
\kappa^2 g_{W^\prime}^2
\frac{|\mathcal{A}_{VV^\prime}|^2 }{128 \pi(1+\delta_{V,V^\prime})} 
\frac{M_\varphi^2}{M_{W^\prime}^2} .
\eea
The coefficients $\mathcal{A}_{VV^\prime}$ are generated by the loops
of $W^\prime$ shown in Fig.~\ref{fig:AVV_loop}. 
Assuming that the $W^\prime$ is the only particle circulating in the loop and neglecting
the $Z$ mass one obtains
\bea\label{AVV:general}
\!\!\!\!\!\!\!\mathcal{A}_{VV^\prime}\!=\!-C_F\frac{ g_{V W^\prime W^\prime} g_{V^\prime W^\prime W^\prime}}{8\pi^2} \mathcal{F}(\tau),
\eea
where $\tau\equiv M_\varphi^2/4M_{W^\prime}^2$ and $\mathcal{F}(\tau)$ is given by the well known loop function for a vector boson~\cite{Djouadi:2005gi}~\footnote{The corresponding
loop functions for scalar or fermion mediators are smaller than the one for vectors.
Thus, scalar or fermion mediators will require larger couplings to achieve the same 
values of $\mathcal{A}_{VV^\prime}$ as compared with vector mediators with the
same quantum numbers.}
\bea
\!\!\!\!\!\!\!\!\!\!\!\mathcal{F}(\tau) &\!\!\!=&\!\!\! \frac{2\tau^2+3\tau+3(2\tau-1)f(\tau)}{\tau^2} , \nn \\
\!\!\!\!\!\!\!\!\!\!\!f(\tau) &\!\!\!=&\!\!\!
\left\{\begin{array}{l c}\arcsin^2{\sqrt{\tau}}&\tau\leq 1\\-\frac 14\left[\log{\frac{1+\sqrt{1-\tau^{-1}}}{1-\sqrt{1-\tau^{-1}}}}-i\pi\right]^2&\tau>1\end{array}\right. \!\!\!.
\eea
The color factors are given by
\bea
C_F= \left\{ \begin{array}{ll}
d_r& V,V^\prime=\gamma,Z\\
C(r)& V=V^\prime=g
\end{array}\right . ,
\eea
where $C(r)$ and $d_r$ are the index and dimension of the color
representation of $\varphi$ respectively. In the limit $M_{W^\prime} \gg M_\varphi/2$
the loop functions quickly converge to constant values giving,
\bea
\mathcal{A}_{VV^\prime} =  - \frac{7}{8\pi^2} 
C_F g_{V W^\prime W^\prime} g_{V^\prime W^\prime
  W^\prime}. 
\eea
Regarding the phenomenological implications of the $W^\prime$, it
can always be pair produced at the LHC. (Single production is also possible, 
but only for very specific quantum numbers of the $W^\prime$.)
 The possible decay channels are very sensitive to
its particular quantum numbers and the eventual presence of extra
particles beyond our simplified model spectrum. Such particles do not
have to play any role in the diphoton anomaly but can be necessary to
mediate the decay of the $W^\prime$ depending on its quantum numbers.
Generically, the new vector will decay to pairs of quarks or to a SM
$W$ plus a number of colored particles (gluons or quarks). Thus, the
collider signature of this new vector boson would be pairs of dijet
or $W+n-\mathrm{jets}$ resonances at the $W^\prime$ mass.

\section{Partial Width to Mass Ratios}

Considering first the $\gamma \gamma$ and $gg$ partial width to mass ratios we obtain
\bea
\!\!\!\!\!\!\!\!\!\!\!\frac{\Gamma(\varphi \to \gamma \gamma)}{M_\varphi} &\!\!\!\!\!\approx&\!\!\!\!\!  
5.3\! \times\! 10^{-8} \kappa^2 g_{W^\prime}^2 Q^4 d_r^2\!
\left(\frac{1\mbox{ TeV}}{M_{W^\prime}}\right)^2\!\!\!,\label{phigaga} \\
\!\!\!\!\!\!\!\!\!\!\!\frac{\Gamma(\varphi \to gg)}{M_\varphi} &\!\!\!\!\!\approx&\!\!\!\!\!  
8.7\! \times\! 10^{-6} \kappa^2 g_{W^\prime}^2 C(r)^2\!
\left(\frac{1\mbox{ TeV}}{M_{W^\prime}}\right)^2\!\!\!\!.
\label{phigg}
\eea
From here we see that in order to reproduce the diphoton excess
we need $\Gamma(\varphi \to \gamma \gamma)/M_\varphi \gtrsim 1.1\times
10^{-6}$, where the lower bound is obtained for $\Gamma_\varphi\approx \Gamma(\varphi \rightarrow gg)$.~This translates into the bound on the new gauge coupling ($Q\not= 0$)
\bea\label{gWpconstr}
g_{W^\prime} \gtrsim 
\frac{4.56}{d_r \kappa Q^2} 
\left(\frac{M_{W^\prime}}{1\mbox{ TeV}}\right),
\eea
as well as the bound
\bea\label{Gammaggconstr}
\frac{\Gamma(\varphi \to gg)}{M_\varphi} \gtrsim 
1.8\times 10^{-4} \left(\frac{C(r)}{d_r Q^2}\right)^2.
\eea
Thus we see that the coupling remains perturbative for a large range
of masses, even with relatively mild values for the parameters
$\kappa$, $Q$ and $r$. As an example, using $\kappa=\sqrt{2}$ we consider 
the case of a color octet representation with $Q=1$, as well as a color triplet of charge $Q=5/3$.
From Eq.~(\ref{gWpconstr}), we obtain in both cases
\be\label{eq_gW_exampl}
g_{W^\prime} \gtrsim 0.4 
\left(\frac{M_{W^\prime}}{1\mbox{ TeV}}\right),
\ee
while from Eq.~(\ref{Gammaggconstr}) we find
\be\label{eq_Gammagg_exampl}
\frac{\Gamma(\varphi \to gg)}{M_\varphi} \gtrsim \left\{\begin{array}{c l}2.5\times 10^{-5}&\mbox{for }d_r=8,~Q=1\\
4.6\times 10^{-6}&\mbox{for }d_r=3,~Q=\frac 53\end{array}\right.,
\ee
which are both compatible with the reported
excess~\cite{Franceschini:2015kwy}.\footnote{Note that the octet
  representation can be coupled to a fermionic current,
  $\overline{d_R^i}T_A \gamma_\mu u_R^j$, and could therefore show up
  in dijet or single top searches. Likewise, 
the color triplet vector field can couple to $\overline{e_R^i}\gamma_\mu u_{R}^j$ and thus 
be produced in $t$-channel in $pp\rightarrow \ell^+ \ell^-$. 
In any case, since our study is largely  
independent of such fermionic couplings, one can avoid
direct limits on $M_{W^\prime}$ 
by considering fermiophobic vector bosons.}

As we have shown in the examples above, only relatively small couplings at the TeV scale are required to explain the diphoton excess, and our perturbative expansion is well under control.
Furthermore, although the running of the different couplings in a theory is a highly model-dependent issue\footnote{Unlike other scenarios proposed to explain the diphoton excess using vectorlike fermions and/or scalars~(see e.g. \cite{Franceschini:2015kwy,Salvio:2015jgu,Son:2015vfl,Goertz:2015nkp,Salvio:2016hnf} for renormalization group studies in such models), our simplified model is non-renormalizable by construction and thus we need to know the details of a minimal renormalizable embedding in order to perform a meaningful study of its behaviour in the ultraviolet.
}, 
note that we are assuming the new vectors arise from the symmetry breaking of an extended gauge sector, 
which must also be non-abelian in order to provide color and/or electric charges to $W^\prime$. The fact that
we are dealing with the coupling of a non-abelian gauge theory generically improves its ultraviolet behaviour. 
Indeed, provided the ultraviolet completion of our simplified model is a non-abelian gauge group, its coupling is likely to
be asymptotically free, unless a large matter content is present in the model.
To illustrate this, we consider an explicit realization of our mechanism~\cite{Blumhofer:1997vb} in which the new vector bosons correspond to our second example in Eqs. (\ref{eq_gW_exampl}) and (\ref{eq_Gammagg_exampl}), {\it i.e.} they are an electroweak singlet, color triplet of hypercharge 5/3. They arise from the spontaneous symmetry breaking of an $SU(4)\times SU(3) \to SU(3)_c$ gauge group. In particular, they are part of the $SU(4)$ gauge bosons, which interact with a coupling $g_4$ that represents our $g_{W^\prime}$ (and a group theory factor of $\kappa=\sqrt{2}$). The one-loop beta function of a general gauge group with arbitrary fermion and scalar content reads \cite{Cheng:1973nv,Machacek:1983tz}
\begin{equation}
\beta(g)=-\frac{g^3}{16\pi^2} \left\{ \frac{11}{3} C_2(G) - \frac{2}{3} S_2(F) -\frac{1}{6} S_2(S) \right \},
\end{equation}
where $C_2(R)$ is the quadratic Casimir of the corresponding representation $R$ and $S_2$ the Dynkin index of the corresponding representations of the fermions ($F$) and scalars ($S$). We have assumed that, as is our case, the fermions correspond to 2-component spinors. In the case at hand we have 3 families of fermions in the $4$ representation of $SU(4)$ and a total of 11 scalars in the $\bar{4}$ representation (see Ref.~\cite{Blumhofer:1997vb} for details). Putting all together we
find
\begin{equation}
\beta(g_4)= -\frac{g_4^3}{16\pi^2} \frac{51}{4} <0,
\end{equation}
and therefore the corresponding gauge coupling runs towards smaller values in the ultraviolet.

Thus, we see that if the new vector boson has both electric and
color charges, the diphoton excess can be explained entirely in terms
of a simplified model consisting of the new vector boson and the
scalar that is responsible for its mass.~Regarding other channels, we
also see that the decay into gluons can have a smaller decay width
than the decay into photons.~Being an electroweak singlet, the
$W^\prime$ boson will induce $\gamma Z$ and $ZZ$ couplings that scale
according to the tangent of the weak angle. Thus, we have the generic
prediction 
\bea
\frac{\Gamma(\varphi \to \gamma Z)}{\Gamma(\varphi \to \gamma \gamma)} &=&
2 (s_W/c_W)^2, \\
\frac{\Gamma(\varphi \to Z Z)}{\Gamma(\varphi \to \gamma \gamma)} &=&
(s_W/c_W)^4, \\
\Gamma(\varphi \to W^+ W^-) &=&0.
\eea
Considering electroweak non-singlets which would give $W^+W^-$ decays 
would also be interesting. As in the case of $SU(2)_L$ singlets, the only generic predictions
for the different decay widths are the ones implied by gauge invariance. These have
been considered in the literature, see Eq.~(2.14) in \cite{Franceschini:2015kwy}. 

Finally, as in most explanations of the diphoton anomaly, the total decay width of the
new resonance will depend on the presence of extra new particles and couplings in the 
explicit model embedding of our simplified scenario. Hence, depending on the ultraviolet completion,
large or small widths could be obtained without affecting the explanation of the diphoton excess.

\section{Link to other anomalies}
\label{section:diboson}

We have seen in the previous section that the vector boson of a
spontaneously broken gauge symmetry, together with the scalar
responsible for its mass, can generate the observed excess in the
diphoton spectrum.~As we argued in the introduction, new vector bosons
are further motivated by other anomalies reported by experimental
collaborations. 

In particular, the LHC experiments have observed several excesses in various
diboson channels at around 2 TeV,~e.g.~\cite{Aad:2015owa,Khachatryan:2014hpa,Khachatryan:2014gha}~(See also~\cite{Dias:2015mhm} for a combination of both ATLAS and CMS results).~While the new LHC data at 13 TeV does not see an excess in the same region, the collected statistics does not provide enough sensitivity to exclude a new resonance with this mass~\cite{ATLASDiboson13TeV1,ATLASDiboson13TeV2,ATLASDiboson13TeV3,ATLASDiboson13TeV4,CMS:2015nmz}.~Several explanations for this excess have been proposed~\cite{Fukano:2015hga,
Hisano:2015gna,
Franzosi:2015zra,
Cheung:2015nha,
Dobrescu:2015qna,
Aguilar-Saavedra:2015rna,
Gao:2015irw,
Thamm:2015csa,
Brehmer:2015cia,
Cao:2015lia,
Cacciapaglia:2015eea,
Abe:2015jra,
Allanach:2015hba,
Abe:2015uaa,
Carmona:2015xaa,
Dobrescu:2015yba,
Chiang:2015lqa,
Cacciapaglia:2015nga,
Sanz:2015zha,
Chen:2015xql,
Omura:2015nwa,
Anchordoqui:2015uea,
Chao:2015eea,
Bian:2015ota,
Kim:2015vba,
Lane:2015fza,
Liew:2015osa,
Arnan:2015csa,
Dev:2015pga,
Coloma:2015una,
Goncalves:2015yua,
Aguilar-Saavedra:2015yza,
Fichet:2015yia,
Petersson:2015rza,
Deppisch:2015cua,
Zheng:2015dua,
Llanes-Estrada:2015hfa,
Aydemir:2015nfa,
Chen:2015cfa,
Bian:2015hda,
Bandyopadhyay:2015fka,
Awasthi:2015ota,
Li:2015yya,
Ko:2015uma,
Collins:2015wua,
Allanach:2015blv,
Dobrescu:2015jvn,
Sajjad:2015urz,
Wang:2015sxe,
Feng:2015rzn,
Das:2015ysz,
Aguilar-Saavedra:2015iew,
Aydemir:2015oob
}
(see~\cite{Brehmer:2015dan} for a recent review) many which invoke the presence of a new massive vector boson resonance and in particular a $W^\prime$ vector boson with electric charge $\pm 1$.~It is therefore interesting to consider the possibility that such a new particle could also have a role in the diphoton anomaly, thus linking both excesses.

If a new vector resonance is indeed responsible for the diboson excess, this implies that it must be a
color singlet.~From Eq.~(\ref{gWpconstr}) we then have (for $Q=\pm1$, $d_r=1$,
$M_{W^\prime}=2\mbox{ TeV}$), 
\be
g_{W^\prime} \gtrsim 9.12/\kappa.
\ee
The group-theory factor $\kappa$ can be larger than one, but not much unless very large group representations
are chosen. For instance, if
the $W^\prime$ comes from an additional $SU(2)$ group
broken by a scalar triplet we have $\kappa=\sqrt{2}$ and relatively large values of the coupling constant satisfying $g_{W^\prime}\gtrsim 6.45$ are needed to generate an adequate $\Gamma(\varphi\rightarrow \gamma\gamma)$ partial width.~Of course, the $W^\prime$ is now a color singlet and so cannot mediate the production of the new scalar particle in gluon fusion.~One could still produce $\varphi$ through $b\bar{b}\rightarrow \varphi$, via $W^\prime$ loops.~In that case, however, according to \cite{Franceschini:2015kwy}, Eq.~(\ref{gaga_gg}) must be replaced by 
\begin{equation}
\frac{\Gamma(\varphi\to \gamma \gamma)}{M_\varphi} 
\frac{\Gamma(\varphi\to b\bar{b})}{M_\varphi} 
\approx 1.9\times 10^{-4} \frac{\Gamma_\varphi}{M_\varphi},
\end{equation} 
which implies large non-perturbative values of the $\varphi W^\prime W^\prime$ coupling $g_{W^\prime}\gtrsim 85$.~Moreover, generating the adequate effective $\varphi b\bar{b}$ coupling would also require large $W^\prime t\bar{b}$ couplings, in conflict with the results of direct $W^\prime$ searches.~Similar conclusions hold for attempting to produce $\varphi$ through photon fusion~\cite{Fichet:2015vvy,Csaki:2015vek} via a $W^\prime$ loop.~It is therefore clear that both excesses are difficult to connect
by means of a single uncolored $W^\prime$ vector of charge $\pm 1$, given that 
other possible production mechanisms will typically induce further activity in the
diphoton events that has not been observed experimentally.

Of course, it is possible that different new particles can mediate the production of $\varphi$.~Extending the simplified $W^\prime$ model with new colored particles could then account for both diphoton and diboson excesses.~One possibility is simply to introduce extra vector-like fermions: quarks to generate the required coupling to gluons, and leptons to make the decay width into
photons arbitrary. (See for instance Refs.~\cite{Franceschini:2015kwy,Falkowski:2015swt} for examples in which this possibility has been explored.)
In this initial exploration however, we find it more interesting to focus 
on vector boson solutions to other experimental anomalies.~One such example
that could produce the required $\varphi gg$ coupling is the case of a
relatively light axigluon, which is motivated by the
$\sim 3~\sigma$ discrepancy in the forward-backward $t\bar{t}$
asymmetry observed at the Tevatron~\cite{Barcelo:2011vk,
Tavares:2011zg,
Alvarez:2011hi,
AguilarSaavedra:2011ci,
Krnjaic:2011ub,
AguilarSaavedra:2011ck,
Aguilar-Saavedra:2014nja
}.~Assuming that both $W^\prime$ and the axigluon $G^\prime$ get all
their mass from $\varphi$ we obtain a relation between their couplings
to $\varphi$
\begin{equation}
\frac{g_{W^\prime} \kappa_{W^\prime}}{M_{W^\prime}}
=\frac{g_{G^\prime} \kappa_{G^\prime}}{M_{G^\prime}}.
\end{equation}
This automatically fixes the decay width into gluons from the quantum
numbers of the axigluon ($C(8)=3$ and $Q=0$), as a function
of the axigluon mass (the dependence on the mass comes exclusively from
the loop function in Eq.~(\ref{AVV:general})). While low axigluon masses are
favored by the discrepancy in the forward-backward $t\bar{t}$ asymmetry, 
they increase the decay of the new scalar into $gg$, which is constrained 
by dijet searches. We find
\bea
\frac{\Gamma(\varphi \to gg)}{M_\varphi} \gtrsim 1.7\times 10^{-3},
\eea
for $M_{G^\prime}\leq 1$ TeV (close to the absolute lower bound
$\Gamma(\varphi \to gg)/M_{\varphi} \gtrsim 1.6\times 10^{-3}$ obtained in
the large axigluon mass limit, $M_{G^\prime}\gg M_{\varphi}/2$). 
This value is compatible with the diphoton excess but shows some
tension with bounds from dijet
searches~\cite{Franceschini:2015kwy}.~This means that, if this is the
origin of the diphoton anomaly, either an excess in dijet searches at
an invariant mass of $\sim 750$ GeV should be seen soon or $\varphi$
is not fully responsible for the mass of the axigluon.

\section{Conclusions}
\label{section:conclusions}

The first results from the LHC Run 2 have provided a tantalizing excess in the 
diphoton channel at invariant mass around 750 GeV. It this persists, it would be
an unambiguous sign of physics beyond the SM.~The decay to photons implies the potential new resonance is likely a scalar particle, with loop-induced decays.~While most solutions addressing this excess thus far have relied on including extra vector-like quarks to generate the necessary effective couplings to gluons and photons, in this letter we have explored the possibility that the new scalar is responsible for the symmetry-breaking of an extended gauge sector.~The couplings to gluons and photons are generated by the corresponding charged massive vector bosons.

We have explored this possibility by employing a simplified model consisting of an electroweak singlet vector boson with arbitrary color and electric charge, and discussed the conditions under which the diphoton excess can be explained.~In this simplified model the decay channels into neutral electroweak gauge bosons are fixed by gauge invariance while there is no decay to $W^+W^-$. Of course, electroweak non-singlet representations could be considered, in which case there will also be decays into $W^+W^-$.

Finally, the diphoton excess joins a list of intriguing hints of
potential new physics that have been observed in recent experiments at
the LHC and Tevatron, such as  the $\sim 2$ TeV 
diboson anomaly, or
the $t\bar{t}$ forward-backward asymmetry anomaly.
We have shown that the three anomalies can in principle be connected. 
The 2 TeV $W^\prime$ that can potentially explain the diboson anomaly
can also be a good mediator to induce the scalar couplings to pairs of
electroweak vector bosons, but this requires a sizeable
$\varphi W^\prime W^\prime$ coupling. Similarly, one of the best candidates to explain
the anomalous $t\bar{t}$ forward-backward asymmetry, a light axigluon,
can induce the needed coupling to gluons that account for the scalar
resonance production, although some tension with dijet searches is found
in this case.

To conclude, we have shown in this letter that the diphoton excess can be
associated to a heavy Higgs mechanism, which predicts the existence of new heavy vector bosons.

\newpage

{\bf Acknowledgments:}
The work of J.B. has been supported by the European Research Council
under the European Union's Seventh Framework Programme
(FP~\!\!/~\!\!2007-2013)~\!/~\!ERC Grant Agreement n. 279972. 
J.S. and R.V.M.~are supported by MINECO, under grant number
FPA2013-47836-C3-2-P. J.S. is also supported by MINECO grant
FPA2010-17915 (Fondos FEDER), the European Commission contract PITN-GA-2012-316704
(HIGGSTOOLS) and by Junta de Andaluc\'{\i}a grants FQM 101 and FQM 6552. 
